# Solitons, compactons and undular bores in Benjamin-Bona-Mahony-like systems

Aparna Saha [1], B. Talukdar [1], Umapada Das [2] and Supriya Chatterjee [3]

[1] Department of Physics, Visva-Bharati University, Santiniketan 731235, India

[2] Department of Physics, Abhedananda College, Sainthia 731234, India

[3] Department of Physics, Bidhannagar College, EB-2, Sector-1, Kolkata 700064, India

We examine the effect of dissipation on traveling waves in nonlinear dispersive systems modeled by Benjamin- Bona- Mahony (BBM)-like equations. In the absence of dissipation the BBM-like equations are found to support soliton and compacton/anticompacton solutions depending on whether the dispersive term is linear or nonlinear. We study the influence of increasing nonlinearity of the medium on the soliton- and compacton dynamics. The dissipative effect is found to convert the solitons either to undular bores or to shock-like waves depending on the degree of nonlinearity of the equations. The anticompacton solutions are also transformed to undular bores by the effect of dissipation. But the compactons tend to vanish due to viscous effects. The local oscillatory structures behind the bores and/or shock-like waves in the case of solitons and anticompactons are found to depend sensitively both on the coefficient of viscosity and solution of the unperturbed problem.

## 1. INTRODUCTION

It is fairly well known that Korteweg-de Vries (KdV) equation

$$u_t + u_x + uu_x + u_{xxx} = 0, u = u(x,t),  \qquad (1)$$

which could model generation of solitons on the surface of water, possesses many remarkable properties. For example, the investigation of conservation laws of the equation led to the discovery of a wide variety of ingenuous mathematical techniques including the Muira transformation, Lax-pair representation, inverse scattering method and Bi-Hamiltonian structure that were subsequently used to examine the integrability of other similar equations. But it is less well known that (1) has an unbounded dispersion relation. This awkward physical constraint can be realized in terms of the linearized form $u_t + u_x + u_{xxx} = 0$ of the KdV equation. Assuming the solution of the linear equation as a summation of Fourier component $f(k)e^{-i(kx-\omega t)}$ we obtain the dispersion relation $\omega(k) = k - k^3$. The corresponding phase velocity $v_p (= \omega(k)/k) = 1 - k^2$ becomes negative for $k^2 > 1$. This contradicts our initial assumption for the forward traveling wave. Moreover, the group velocity $v_g (= \dfrac{d\omega}{dk}) = 1 - 3k^2$ has no lower bound.

To circumvent the above difficulties originally Peregrine [1] and subsequently Benjamin, Bona and Mahony [2] proposed the equation

$$u_t + u_x + uu_x - u_{xxt} = 0 \qquad (2)$$

as an alternative model for the motion of long waves in nonlinear dispersive systems. The linearized version of (2) leads to the dispersion relation $\omega(k) = k/(1+k^2)$ such that both $v_p$ and $v_g$ are well behaved for all values of $k$. Thus, as opposed to the KdV equation, (2) provides us with a regularized long wave (RLW) equation. The authors of ref. 2 established that both (1) and (2) are valid at the same level of approximation but in applicative context the latter equation does have some advantage over the KdV equation. As a result (2) is often called the Benjamin-Bona-Mahony or BBM equation.

In order to understand the role of nonlinear dispersion in the formation of patterns in waves governed by the RLW model, Yandong [3] and Wang et al [4] considered a family of BBM-like equations

$$u_t + u_x + a(u^m)_x - (u^n)_{xxt} = 0, m \geq 2,$$
$$n \geq 1. \qquad (3)$$

For $m = 2, a = 1/m$ and $n = 1$, (3) reduces to the usual BBM equation. The equation is more nonlinear than the usual BBM equation for $m > 2$. For values of $n > 1$ the dispersive term in (3) is nonlinear. Evolution equations with nonlinear dispersive terms were first considered by Rosenau and Hyman [5]. In particular, the equation

$$u_t + (u^m)_x + (u^n)_{xxx} = 0, \; m > 1, \; 1 < n \leq 3 \qquad (4)$$

was used by them as a model that was expected to account for the formation of patterns in liquids. It was demonstrated that the traveling wave solutions of (4) are free from the usual exponential tails of solitons and vanish identically outside a finite range. These solutions were given the name compactons. The compactons are robust within their range of existence. However, unlike the interaction of solitons in the KdV-like systems, the point at which two compactons collide is marked by the birth of a low amplitude compacton-anticompacton pair.

A straightforward generalization to the family of equations given in (3) to include the effect of viscosity is provided by

$$u_t + u_x + a(u^m)_x - (u^n)_{xxt} = \eta u_{xx}, \qquad (5)$$

where $\eta$ is the kinematic viscosity coefficient. In the recent past Mancas et al. [6] derived a formalism to write the general solution of (5) for $n=1$, $m=2$ and $a=1/m$, and demonstrated that in the case of BBM equation there still exists, in certain region of space, bounded traveling wave solutions in the form of solitons. The results of these authors appear to substantiate a general remarks made by El et al. [7] who claimed that introduction of small dissipation in a nonlinear system dramatically changes its properties, allowing in some cases for the presence of steady solution.

In this paper we shall first construct analytical solutions of (3) for $n=1$ and $2$, and in each case we shall consider different values of $m$ with a view to illustrate how the solutions behave as the systems become more and more nonlinear. We then turn our attention to study the effect of dissipation on the system modeled by (3). We shall achieve this by solving (5) and comparing its solutions with the appropriate solutions of (3). It appears that the mathematical approach developed in ref. 6 which was used to solve the initial boundary value problem for the dissipative BBM equation (m=2 and n=1) is not applicable to the general equation given in (5). In view of this, we convert the equation to a Cauchy problem in an appropriate coordinate system and solve it by using numerical methods. Interestingly, we find that solutions of the dissipative system, in general, do not represent bounded traveling waves. On the other hand, the solution obtained by us for any chosen set of values for $n$, $m$ and $\eta$ resembles either the so-called undular bores or shock waves. As with shock –like waves, the bore is a well known phenomenon in fluid mechanics, describing the transition between two uniform streams with different flow depths [8]. Undular bores feature free surface oscillations behind the front of the bore, and one says that the bore is purely undular if none of the waves behind the bore are breaking [9].

In sec. 2 we introduce the so-called traveling coordinates and make use of it to convert the partial differential equation in (3) to a nonlinear ordinary differential equation and subsequently write the latter as a dynamical system. We provide necessary phase-plane analysis to study the behavior of the nonlinear system. We also present plots of the vector fields which play a role to model the speed and direction of a moving fluid throughout the space. All results are presented with a view to examine the effects of varying nonlinearity and dispersion (different values of m and n) on the soliton- and compacton solutions in the non-dissipative medium. In sec. 3 we examine the effects of dissipation on the wave motion in BBM-like systems. As opposed to the case studies presented in sec. 2, here we take recourse to the use of numerical routines to deal with equations in (5). We find that due to viscous effects the solitons and compactons are converted either into undular bores or shock-like waves the exact nature of which, depends on the values of m and n. Finally in sec.4 we summarize our outlook on the present problem and make some concluding remarks.

Equation (3) which models a family of non-dissipative BBM-like systems for different values of $m$ and $n$ does not involve the space and time coordinates explicitly. Consequently, the equation is invariant under translation in these variables and can, therefore, be reduced to a nonlinear ordinary differential equation using the traveling coordinates $\xi = x - vt$ with $v$, the nonzero translational wave velocity. Keeping this in mind we apply the change of variable

$$u(x,t) = \phi(x - vt) \qquad (6)$$

in (3) and obtain the ordinary differential equation

$$(1-v)\phi' + a(\phi^m)' + v(\phi^n)''' = 0, \qquad (7)$$

where primes denote differentiation with respect to $\xi$. Integration of (7) with respect to $\xi$ yields

$$(1-v)\phi + a(\phi^m) + v(\phi^n)'' = 0. \qquad (8)$$

In writing (8), as in ref.2, we have taken the constant of integration equal to zero. This choice will allow us to reproduce, in the appropriate limit, all results of the BBM equation from the solutions of (8). We now make a further change in the dependent variable of (8) by writing

**2.NONDISSIPATIVE BBM-LIKE SYSTEMS**

$\phi = y^{\frac{1}{n}}$ and express the transformed equation in the form

$$y'' + \frac{a}{v} y^{\frac{m}{n}} + \frac{1-v}{v} y^{\frac{1}{n}} = 0. \qquad (9)$$

To treat (9) by the methods of dynamical systems theory [10] we rewrite the equation as a system of first-order differential equations

$$z = y', \; z = z(\xi) \qquad (10a)$$

and

$$z' = -\frac{a}{v} y^{\frac{m}{n}} - \frac{1-v}{v} y^{\frac{1}{n}}. \qquad (10b)$$

On the other hand, we could multiply (9) by $y'$ and integrate once to get

$$c = \frac{n(1-v)}{n+1} y^{\frac{n+1}{n}} + \frac{an}{m+n} y^{\frac{m+n}{n}} + \frac{v}{2} y'^2, \qquad (11)$$

where $c$ is a constant of integration. Since (9) is an autonomous differential equation without explicit time dependence, the constant $c$ can be identified with the Hamiltonian of the system. The set of equations as given in (10) can be used to draw the phase portrait while (11) can be used to plot the vector fields. The first-order equation in (11) can easily be integrated to write $y$ as a function of $\xi$ for all values of $c$. This gives the solution of the BBM-like equation for given values of $m$ and $n$.

In terms of $\phi$, (10a) and (10b) read

$$z = n\phi^{n-1}\psi, \; \psi = \phi' \qquad (12a)$$

and

$$z' = -\frac{a}{v}\phi^m - \frac{1-v}{v}\phi. \qquad (12b)$$

It is straightforward to combine (12a) and (12b) to write

$$z = \pm\sqrt{\frac{2n\phi^n}{v}\left(\frac{v-1}{n+1}\phi - \frac{a}{m+n}\phi^m\right)}. \qquad (13)$$

Equations (12a) and (13) can now be used to plot the phase trajectories on the $(\phi, \psi)$ plane. In terms of $\phi$ the constant $c$ in (11) can be written as

$$c = \frac{n}{n+1}(1-v)\phi^{n+1} + \frac{an}{m+n}\phi^{m+n} + \frac{vn^2}{2}\phi^{2(n-1)}\psi^2. \qquad (14)$$

In the following we use the above equations to present the phase portrait, vector plot and traveling wave solution of (3) for specific values of $n$ and $m$. We divide our computed results into two distinct classes depending on the values of $n$ (1 or 2). Independently of the values of $m$, the dispersive term in each equation for $n = 1$ is linear while such a dispersive term is nonlinear for any equation with $n=2$. We shall present all numerical results for $v=1.5$.

**(i) Equations with linear dispersive term (n=1)**

Here we begin with $m=2$ and then consider equations for higher values of $m$. We shall see that, for all values of $m$, the equations support soliton solutions although their phase trajectories and vector fields are quite different. As m increases we, however, observe a regularity for changes in the phase-space structure and associated vector field.

(a) $m=2$: In this case, (3) gives the BBM equation. The associated phase portrait and vector field are given in FIGS 1 and 2. We calculated the phase trajectory using (12a) and (13). The vector field was generated from the integral curve in (14).

A phase path that separates obvious distinct regions in the phase plane is known as the separatrix. The phase path in FIG.1 joins the saddle-type equilibrium (0,0) to itself by enclosing a center-type equilibrium point at (1.5,0) and is thus a form of separatrix often known as the homoclinic path. From FIG.2 we see that at all points the vector field diverges from the critical point (0,0) while similar tangent vectors converge towards (1.5,0). This reconfirms that the equilibrium point (0,0) is a saddle and (1.5,0) is a center.

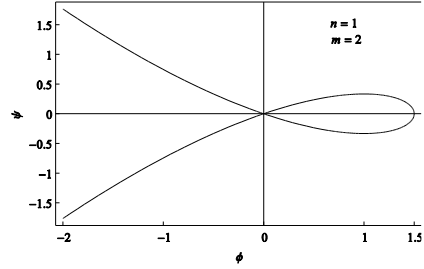

FIG.1. Phase diagram of the BBM equation:

Eq.(3) : $n=1, m=2$.

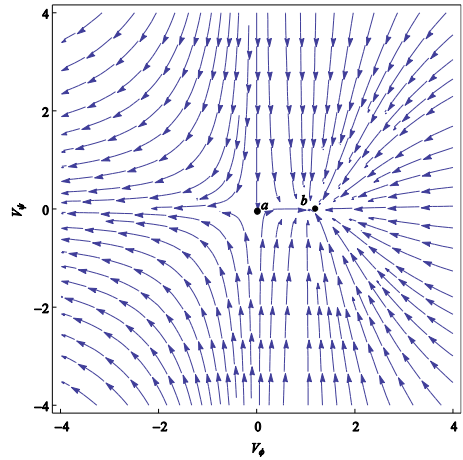

FIG.2.(Color online):Vector flow for the integral Curve (14) appropriate for the BBM equation. The coordinates $V_\phi = -\dfrac{\partial c}{\partial \phi}$ and $V_\psi = -\dfrac{\partial c}{\partial \psi}$.

The orientation of the line segment indicates the slope $\frac{d\psi}{d\phi}$ at the position of the center of the segment.

For $c = 0, n = 1, m = 2$ and $v = 1.5$ we obtained from (11) the well known soliton solution of the BBM equation as

$$\phi(\xi) = 1.5 \mathrm{sech}^2(0.28867\xi). \quad (15)$$

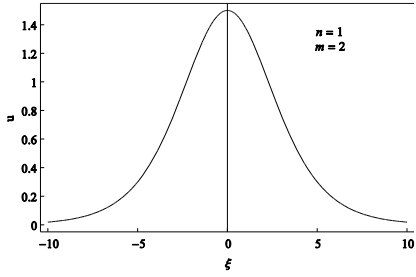

FIG.3. $u(=\phi(\xi))$ in (15) as a function of $\xi$.

The soliton of FIG.3 has an amplitude 1.5 and moves to the right with speed 1.5. Initially, it is centered at the point $x = 0$.

(b) $m = 3$: For this value of $m$, (3) leads to an equation which is more nonlinear than the BBM equation. The present nonlinear equation appears to be the RLW analog of the modified KdV equation. We display the phase diagram and vector field for the equation in FIGS. 4 and 5.

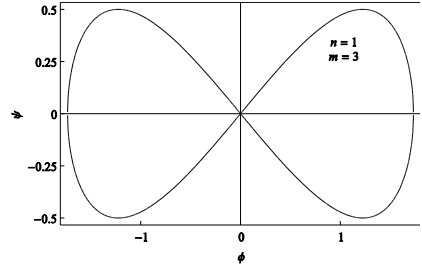

FIG.4. Phase diagram of Eq. (3): $n = 1, m = 3$.

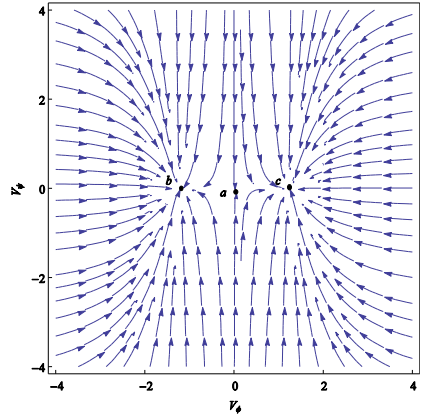

FIG.5. (Color online) Vector flow for the integral curve in (14): $n = 1, m = 3$.

The quantities $V_\phi, V_\psi$ and $\frac{d\psi}{d\phi}$ carry similar meaning as in FIG.2.

The phase trajectory in FIG.4 is a closed homoclinic path that joins the saddle-type equilibrium point $(0,0)$ to itself and encircles two other center-type equilibrium points $(\pm 1.73205)$. As expected the tangent vectors in FIG.5 diverge away from the point $(0,0)$ and converge towards each of the points

($\pm 1.73205$). The solution of the BBM-like equation for $m = 3$ is given by

$$u(x,t) = 1.73 \operatorname{sech}(0.57735\xi). \quad (16)$$

From (15) and (16) we see that the soliton solution found for m=3 depends linearly on the sech function while that for m=2 with quadratic dependence on sech function closely resembles the KdV soliton. Moreover, the argument of the sech function in (16) is exactly twice the argument of the sech function that appears in (15). In FIG.6 we plot $u$ of (16) as a function of $\xi$. This figure clearly shows that the present exponentially localized soliton is taller than the soliton of FIG.3.

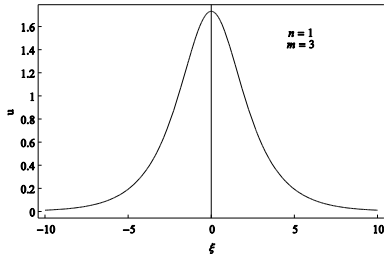

FIG.6. $u$ of (16) as a function of $\xi$.

(c) $m = 4$: In the present case the equation is still more nonlinear. The associated phase diagram and vector field are shown in FIGS. 7 and 8.

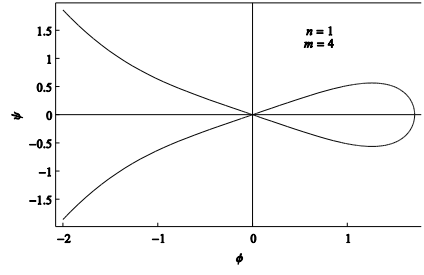

FIG.7. Phase diagram of Eq.(3) $n = 1, m = 4$.

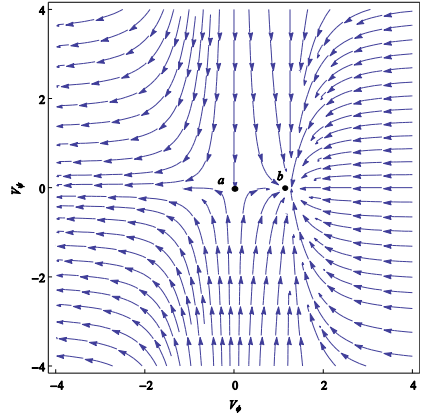

FIG.8. (Color online) Vector flow for the integral curve in Eq.(14): $n = 1, m = 4$. The quantities $V_\phi, V_\psi$ and $\dfrac{d\psi}{d\phi}$ carry similar meaning as in FIG.2.

The phase path in FIG.7 is similar to that in FIG.1 with a saddle point (0,0) and center (1.71,0). As expected, the vector field diverges from (0,0) and converges towards (1.71,0). The solution of (3) for $m = 4$ is found as

$$u(x,t) = 1.71 \operatorname{sech}^{\frac{2}{3}}(0.86602\xi). \quad (17)$$

Note that $u(x,t)$ now has $\operatorname{sech}^{\frac{2}{3}}$ dependence. FIG.9 gives the plot of $u$ in (17) as a function of

$\xi$.

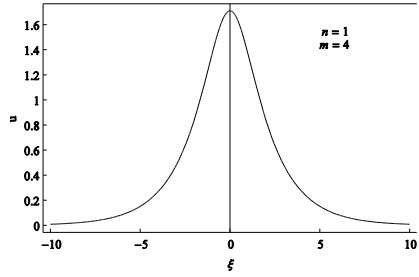

FIG.9. $u$ of (17) as a function of $\xi$

The height of the soliton in the figure is greater than that of the soliton in FIG.3 but less than the height of the soliton in FIG.6.

The phase diagram and vector field of (3) for $m=4$ are identical to those of the same equation for $m=2$. We found similar agreement between the results of equations for $m=5$ and $m=3$. In fact, by considering the phase diagrams and vector plots for still higher values of m we arrived at a conclusion that, so far as the dynamical behavior is concerned, the BBM-like eqations in (3) with linear dipersive term can be divided into two distinct classes depending on whether $m$ is even or odd. All equations with even values of $m$ are characterized by two equilibrium points of which one is a saddle and the other is a center. But equations with odd values of $m$ possess three equilibrium points – one is of saddle type and two others are centers.

The center-type equilibrium point is a sink or attracting fixed point. The saddle-type equilibrium point is a source or repelling fixed point. Thus from the above we infer that equations with even $m$ values physically refer to fluid motion characterized by one source and one sink. On the other hand, equations with odd $m$ values describe motion of fluids in which there are one source point and two sinks.

The solution of (3) for n=1 and an arbitrary value of m can be written in the form

$$u(x,t) = \left[\frac{m(m+1)(v-1)}{2}\right]^{\frac{1}{m-1}} \times \mathrm{sech}\left[\frac{1}{2}(m-1)\sqrt{\frac{v-1}{v}}\xi\right]^{\frac{2}{m-1}}.$$

We made use of the above general expresion to compute $u(x,t)$ as a function of $\xi$ for a large number of values of m>4. From these results and plots in FIGS.3,6 and 9 we found that width of the soliton continuously decreases as m increases. But the amplitude of the soliton first increases from 1.5 to 1.73 as m goes from 2 to 3, then the amplitude decreases and tends to 1 as m increases.

**(ii) Equations with nonlinear dispersive term (n=2)**

For values of $n \geq 2$ the dispersive terms of all equations obtained from (3) are nonlinear. For illustrative purposes we shall consider the case $n = 2$ only and vary the values of $m$ in order to study the effect of higher nonlinearities on the compacton solution supported by the BBM-like equations with nonlinear dispersive terms.

(a) $m = 2$: Here the equation obtained from (8) is given by

$$\phi'' = \frac{v-1}{2v} - \frac{\phi}{4v} - \frac{\phi'^2}{\phi}. \quad (18)$$

By wtiting (18) as two first-order differential equations

$$\psi = \phi' \quad (19a)$$

and

$$\psi' = \frac{v-1}{2v} - \frac{\phi}{4v} - \frac{\phi'^2}{\phi} \quad (19b)$$

we find that (18) has only one equilibrium point $E_1(2(v-1),0)$. Linear stability analysis [10] can now be used to show that $E_1(2(v-1),0)$ is a center-type equlibrium point. Note that for the linear dispersive equation corresponding to that in (18) had two eqilibrium points – one center and other saddle (FIG.1). We display the phase diagram for (18) in FIG.10. The phase trajectory shown for $v = 1.5$ is a closed orbit about the equilibrium point and does not approach $E_1(1,0)$ as $t \to \pm\infty$. The perturbation of the system neither decays to zero nor diverges to infinity but it varies periodically with time. As a result such center-type equilibrium points are often referred to as neutrally stable. Since the center always serves as a sink, the vector field in this case is always directed towards the equilibrium point of (18). We shall not present here the plot of the vector field. In future also, we shall not include any plot of such fields rather we shall assume that these fields converge towards the center and diverge away from the saddle. We found the solution of (18) in the form

$$u(x,t) = 1.333333\sin^2(0.14433\xi). \quad (20)$$

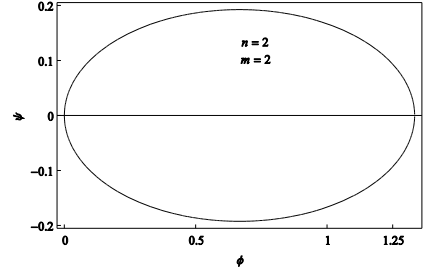

FIG.10. Phase diagram of Eq.(18) $n = 2, m = 2$

The trigonometric solution in (20) for $|\xi| \leq 2\pi$ is shown in FIG. 11. The displayed solitary wave pattern without any exponential tail appears to complement the well known compacton solution of (4) for $m = 2$ and $n = 2$ [5]. Thus we have here an anti-compacton solution.

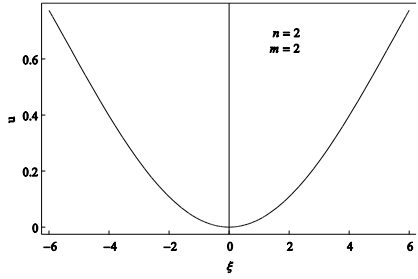

FIG.11 $u$ of (20) as a function of $\xi$

(b) $m = 3$: For this value of $m$, the equation similar to that in (18) reads

$$\phi'' = \frac{v-1}{2v} - \frac{\phi^2}{6v} - \frac{\phi'^2}{\phi}. \tag{21}$$

Equation (21) has two equilibrium points given by $E_1(\sqrt{3(v-1)}, 0)$ and $E_2(-\sqrt{3(v-1)}, 0)$. The equilibrium point $E_1$ is a center while $E_2$ is a saddle. The phase diagram for (21) is shown in FIG.12.

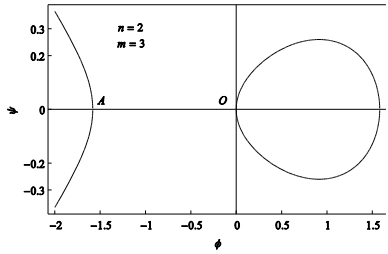

FIG.12. Phase diagram of Eq.(18) $n = 2, m = 3$

The phase trajectory consists of two disjoined curves. The center $E_1$ lies inside the closed elliptical curve. The saddle $E_2$ is located on the $\phi$ axis at a point in between O and A as shown in the figure. Understandably, the vector field will converge towards $E_1$ and diverge away from $E_2$. The solution of (3) for $n = 2$ and $m = 3$ is given by

$$\phi(\xi) = -1.58114 cn^2(-0.18745\xi, 0.5), \tag{22}$$

where $cn(.)$ stands for the Jacobi elliptic cosine function [11]. The plot of $u$ in (22) as a function of the travelling coordinate $\xi$ is shown in FIG.13.

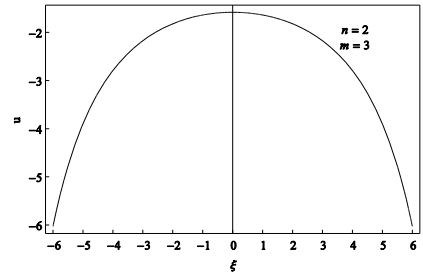

FIG.13. $u$ of (22) as a function of $\xi$

The soliton solution presented above has a compact support and is therefore a compacton. Interestingly, we note that the compacton here appears as an internal wave. Moreover, rather than the trigonometric compacton solutions of Rosenau and Hymann [5], the solution in (22) is given in terms of the Jacobi elliptic function.

(c ) m=4: The BBM-like equation for m=4 is given by

$$\phi'' = \frac{v-1}{2v} - \frac{\phi^3}{8v} - \frac{\phi'^2}{\phi} \quad . \tag{23}$$

As in the case of m=2, (23) has only one equilibrium point $E_1((2(v-1))^{\frac{1}{3}},0)$ and this is a center. The phase diagram for (23) is given in FIG. 14.

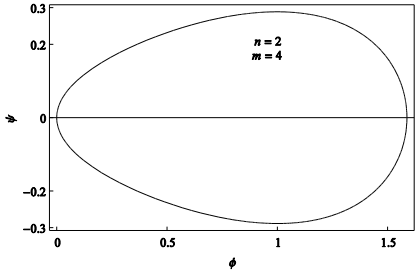

FIG.14. Phase diagram of Eq.(23), n=2 and m=4

As expected, the phase trajectory is a closed path and the center-type equilibrium point lies inside it. The vector field will always be directed towards the center. We obtained the solution of (23) for $v = 1.5$ in terms of complex valued Jacobi sine and cosine functions. The result is given by

$$\phi(\xi) = \frac{a - b\,sn^2(y,m')}{c - d\,cn^2(y,m')} \tag{24}$$

with $a = 5.90557, b = 2.95279 - 5.11438i$

(25a)

$c = 3.22185, d = 1.86014,$
$m' = 0.5 - 0.86602\,5i$

(25b)

and

$y = 1.13005\,7(1+i) -$
$(0.045059 - 1.68163i)\xi$ .

(26)

It may appear from (24) – (26) that $\phi$ in (24) represents a complex solution of (23). However, this is not the case. To substantiate our claim we quote below the results of $u$ for three typical values of $\xi$.

$$u\big|_{\xi=0} = -7.94848 \times 10^{-17} + 6.89183 \times 10^{-16}i,$$

(27a)

$$u\big|_{\xi=5} = 0.68229 + 2.64838 \times 10^{-8}i, \tag{27b}$$

and

$$u\big|_{\xi=10} = 1.53262 - 1.30542 \times 10^{-8}i . \tag{27c}$$

The result in (27a) shows that both real and imaginary parts of $u$ at $\xi = 0$ are zero while the other two results in (27b) and (27c) indicate that imaginary parts of $u$ for $\xi > 0$ are roughly eight orders of magnitude smaller than the corresponding real parts. We have verified that $u$ is an even function i.e. $u(\xi) = u(-\xi)$. Thus as regards (24) the plot of real part of $u$ as a function of $\xi$ will effectively give the variation of $u$ with $\xi$. In FIG. 15 we display the plot of $u$ from (24) as a function of $\xi$.

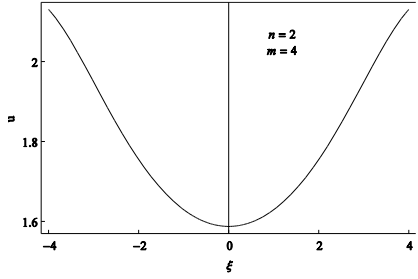

FIG.15. $u$ of (24) as a function of $\xi$

The curve in FIG.15 $(m=4)$ closely resembles to that in FIG.11 $(m=2)$. We have verified that for $n=2$ all equations with even values of $m$ support anti-compacton solutions. In contrast, all such equations for odd values of $m$ model internal waves in the form of compacton solutions.

## 3. DISSIPATIVE BBM-LIKE SYSTEMS

Equations of dissipative BBM-like systems for different values of $m$ and $n$ are given in (5). We write this equation in the traveling coordinate and integrate it once to get

$$(1-v)\phi + a\phi^m + v(\phi^n)'' - \eta\phi' = 0. \quad (28)$$

As before we have taken the constant of integration as zero. The last term in (28) does not permit one to integrate the equation analytically. As a result we shall numerically integrate the eqivalent first-order equations

$$\psi = \phi' \quad (29a)$$

and

$$\psi' = \frac{(v-1)\phi^{2-n}}{vn} - \frac{a\phi^{m-n+1}}{vn} - \frac{(n-1)\psi^2}{\phi} + \frac{\eta\psi\phi^{1-n}}{vn}$$

(29b)

to compute the results for $\phi$ and $\psi$ as a function of $\xi$. Admittedly, the parametric plot of $\psi$ versus $\phi$ will give the phase diagram of the dissipative system. On the other hand, the plot of $\phi$ as a function of $\xi$ will display the solution of the equation.

We regard (29a) and (29b) to define a Cauchy boundary value problem such that these equations could be solved by using prescribed values for $\phi(0)$ and $\phi'(0)$. To solve the dissipative equation for a given set of values for $m$ and $n$ we have chosen to work with the value of $\phi(0)$ taken from the corresponding problem for $\eta = 0$ (Eq. 3) as solved earlier by analytical methods. The value of $\phi'(0)$ is always fixed at zero. We solved the initial value problem given in (29a) and (29b) by taking recourse to the use of fourth-order Runge-Kutta method [12] with an appropriate stability check.

**(i) Dissipative equations with linear dispersive term ( n=1)**

As in the non-dissipative case here we shall present results for $m = 2, 3$ and $4$ with a view to visualize how the nonlinearity of the system affects the dissipative wave. In addition, we shall present for each equation two different sets of results for phase diagram and solution of the equation corresponding to $\eta = 0.1$ and $\eta = 0.5$ with a view to see how the waves in the dissipative medium behave for small and relatively large values for the coefficient of viscosity.

(a) m=2 : For $\eta = 0.1$ the phase diagram and the corresponding solution of the system of equations in (29a) and (29b) are given in FIGS.16 and 17.

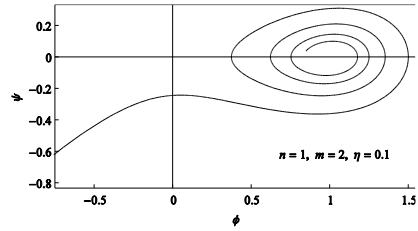

FIG.16. Phase diagram for the dissipative equation with n=1 and m=2 for $\eta = 0.1$

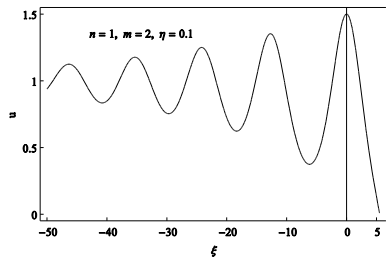

FIG.17. Solution $u$ of the dissipative equation with n=1 and m=2 as a function of $\xi$.

Comparing the curve in FIG.16 with that in FIG.1 we see that due to the effect of dissipation a center type equilibrium has been converted into an unstable spiral. The phase trajectory appears to diverge from the saddle type equilibrium point $(0,0)$ as ocurred in the corresponding nondissipative problem. Similar comparison between the curves of FIGS.3 and 17 indicates that dissipative effect has changed the soliton solution to a undular bore with a local oscillatory structure behibd the bore. These oscillations have their dynamical origin in the periodic solution of the unperturbed system.

In order to see how the formation of the undular bore depends on the viscous effect we portray in FIGS. 18 and 19 the phase portrait and solution of the dissipative system for $\eta = 0.5$ and compare them with the curves in FIGS. 16 and 17.

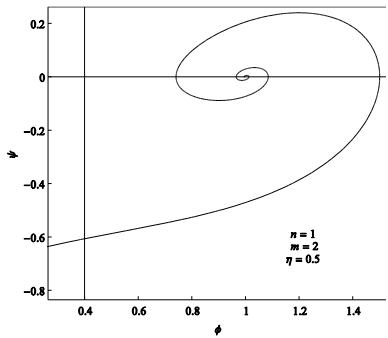

FIG.18. Same as FIG.16 but for $\eta = 0.5$

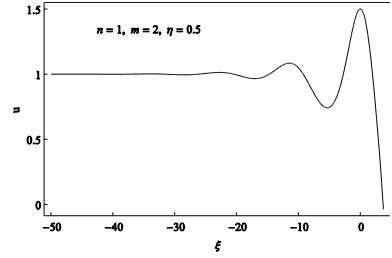

FIG.19. Same as FIG.17 but for $\eta = 0.5$

Closely looking into the curves of FIGS.16 and 18 we see that for $\eta = 0.5$ the spiraling curve leaves the stable point rather quickly than it did for $\eta = 0.1$. Understandably, the observed change in the phase diagram is likely to have some effect on the dynamics of the bore formation. By comparing the curves in FIGS.17 and 19 we confirm that this is indeed the case. Here the dissipative force reduces the amplitudes of the periodic waves following the bore and, in fact, the bore is formed due to ripples in the unperturbed system.

(b) m=3 : In this case, for $\eta = 0.1$ the phase diagram and solution of the associated BBM-like equation are shown in FIGS. 20 and 21.

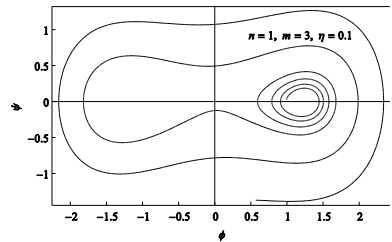

FIG.20. Phase diagram for the dissipative equation with n=1 and m=3 for $\eta = 0.1$

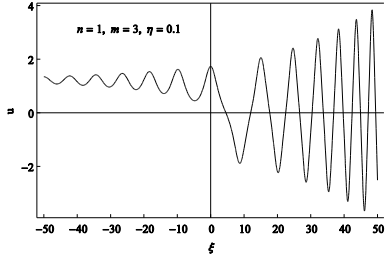

FIG.21. Solution $u$ of the dissipative equation with n=1 and m=3 as a function of $\xi$

We now compare the curves in FIGS. 20 and 21 with those in FIGS.16 and 17 to realize how the nonlinearity of the medium modifies the properties of the dissipative system. The phase trajectories in FIGS.16 and 20 clearly show that as the system becomes more nonlinear the phase path spirals round one of the center type eqilibrium points of the system and encircles two other equilibrium points – one center and other saddle before it leaves them. The observed change in the phase trajectory appears to have some radical effect on the wave of the unperturbed system. For example, instead of a bore formation at $\xi = 0$ (FIG.17), here the wave gains energy from the medium and creates large changes in the medium over very short times (FIG.21). These violent changes cause self steepening of the wave which ultimately resembles a shock front.

For $\eta = 0.5$, the curves corresponding to those in FIGS. 20 and 21 are presented in FIGS. 22 and 23.

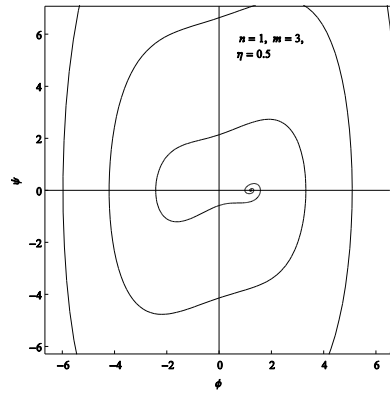

FIG.22. Same as that in FIG. 20 but for $\eta = 0.5$

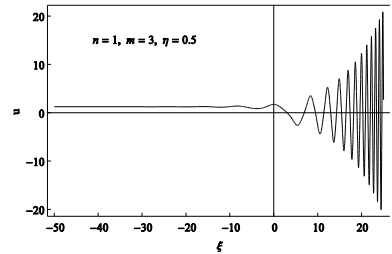

FIG.23. Same as that in FIG. 21 but for $\eta = 0.5$

As is typical for a highly viscous medium, the phase path in FIG.22 closely resembles that in FIGS. 18 for $m = 2$. The plot of FIG. 23 shows that for $\eta = 0.5$ the shock-like behavior of the wave becomes more pronounced.

(c) $m = 4$: Here the phase diagrams and solutions of the dissipative BBM-like equations for $\eta = 0.1$ and $\eta = 0.5$ are displayed in FIGS. 24,25,26 and 27 respectively.

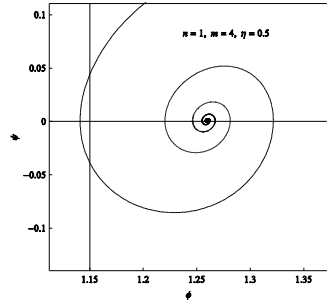

FIG.26. Same as in FIG.24 but for $\eta = 0.5$

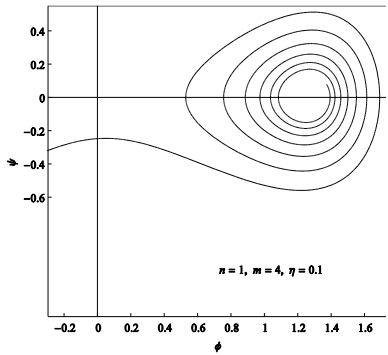

FIG.24. Phase diagram for the dissipative equation with n=1 and m=4 for $\eta = 0.1$

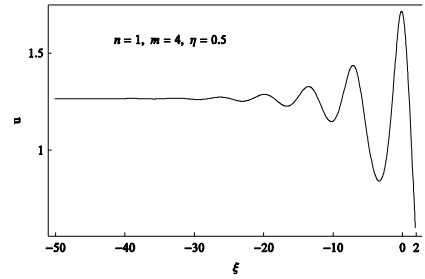

FIG.27. Solution $u$ of the dissipative equation for n=1, m=4 and $\eta = 0.5$ as a function of $\xi$

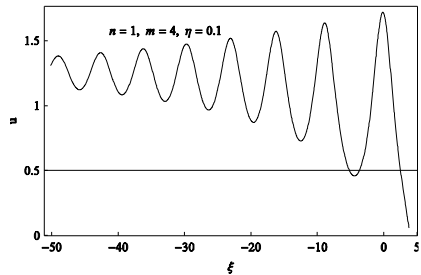

FIG.25 Solution $u$ of the dissipative equation for n=1, m=4 and $\eta = 0.1$ as a function of $\xi$

The plots of FIGS.24 -27 closely resemble those in FIGS. 16 – 19. It thus appears that, as in the case of non-dissipative systems, the phase portraits and solutions of the associated BBM-like equation for m=2 are repeated for m=4 even in the presence of dissipation. Similarly, the phase diagram and solution of equations for m=5 are replicas of those for m=3. In general, we found that the dynamics of all even m equations are identical. The same is also true for all odd m equations.

(ii) **Dissipative equations with nonlinear dispersive term (n=2)**

We have seen that non-dissipative generalized BBM equations with nonlinear dispersive terms support soliton-like solutions with compact support. In particular, the solutions of equations with even m are anti-compactons and those for odd m are compactons. The anti-compacton solutions appear in the form of surface waves while the compacton solutions appear as internal waves. It, therefore, remains an interesting curiosity to examine the effect of dissipation on these robust objects. We shall achieve this by solving the coupled differential equations (29a) and (29b) for n=2 again by using the algorithms of the fourth-order Runge-Kutta method.

(a) m=2: In this case the phase diagram and plot of $u$ as a function of $\xi$ for $\eta = 0.1$ are shown in FIGS.28 and 29.

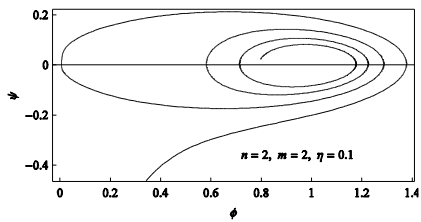

FIG.28. Phase diagram for the dissipative equation with n=2 and m=2 for $\eta = 0.1$

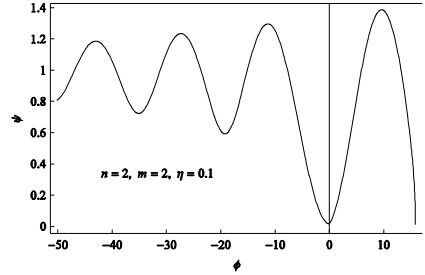

FIG.29. Solution $u$ of the dissipative equation with n=2 and m=2 for $\eta = 0.1$ as a function of $\xi$

In the non-dissipative case, the phase diagram and plot of $u$ as a function $\xi$ for n=2 and m=2 were presented in FIGS. 10 and 11. The corresponding plots for the dissipative case are displayed in FIGS. 28 and 29. Curves in these figures were drawn by solving the initial value problem using $\phi(0) = 0.01$ and $\phi'(0) = 10^{-6}$. Comparing the curves in FIGS.10 and 11 with those in FIGS. 28 and 29 we see that due to the effect of dissipation the center-type equilibrium point has been transformed to a spiral and the anti-compacton to an undular bore. As in FIG.16 (n=1, m=2 and $\eta = 0.1$), the spiral here also corresponds to an unstable focus. But the phase trajectories in these two cases are somewhat different. We also observe a similar difference between the undular bores of FIG.17 (n=1, m=2 and $\eta = 0.1$) and of FIG.29. For example, the bore in FIG.17 appears at $\xi = 0$ while that in FIG.29 appears at a point $\xi > 0$. Understandably, the observed changes in the phase trajectory and corresponding solution of the dynamical equation may be attributed to the nonlinearity of the dispersive term.

We display in FIGS.30 and 31 the phase diagram and $u$ as a function of $\xi$ for $\eta = 0.5$. Comparing the curves of these figures with the corresponding curves of FIGS.28 and 29, we see that for $\eta = 0.5$ the phase trajectory tends to leave the focus more rapidly than it did in FIG.28.

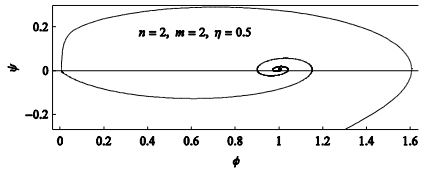

FIG.30. Phase diagram for the dissipative equation with n=2 and m=2 for $\eta = 0.5$

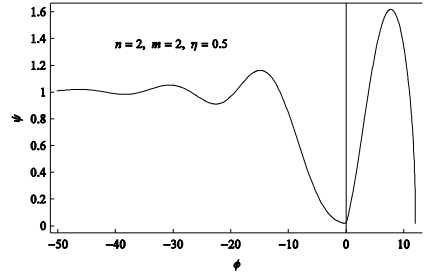

FIG.31 Solution $u$ of the dissipative equation with n=2 and m=2 for $\eta = 0.5$ as a function of $\xi$

As with the result in FIG.29 the bore in FIG.31 is formed at $\xi > 0$. However, the surface oscillation behind the bore appears to be extremely weak. Moreover, the curves in FIGS.30 and 31 are identical to the corresponding curves for n=1, m=2 and $\eta = 0.5$ implying that wave propagation in highly viscous fluid is insensitive to the nonlinearity of the dispersive term in (5).

(b) m=3: Here we solved the coupled differential equations with the initial conditions $\phi(0) = -1.58114$ and $\phi'(0) = 0$ both for $\eta = 0.1$ and $\eta = 0.5$. The appropriate phase portraits and the plot of $u$ as a function of $\xi$ are shown in FIGS. 32 – 35.

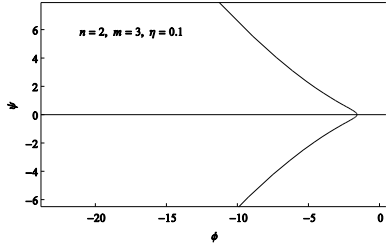

FIG.32. Phase diagram for the dissipative equation with n=2 and m=3 for $\eta = 0.1$

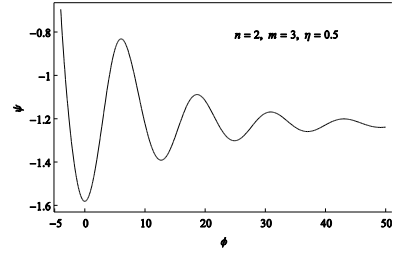

FIG.35 Solution $u$ of the dissipative equation with n=2 and m=3 for $\eta = 0.5$ as a function of $\xi$

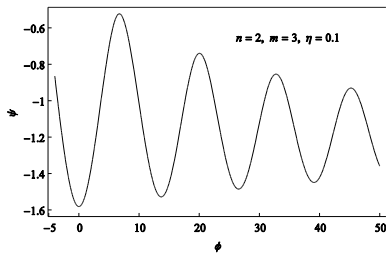

FIG.33. Solution $u$ of the dissipative equation with n=2 and m=3 for $\eta = 0.1$ as a function of $\xi$

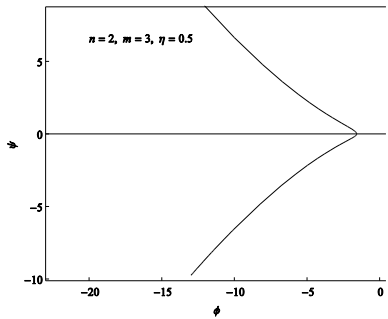

FIG.34. Phase diagram for the dissipative equation with n=2 and m=3 for $\eta = 0.5$

The phase diagram in FIG.32 when compared with that in FIG. 12 shows that the dissipative equation is characterized by only one saddle-type equilibrium point. From the plots in FIGS.13 and 33 we infer that the dissipative effect has converted a compacton into a decaying internal oscillatory wave which tends to disappear for $\xi >> 0$. From FIG.29 we see that due to viscous effect the anticompacton appearing in the form of a surface wave absorbs energy form the medium to culminate in a unimodular bore. On the other hand, the curve in FIG. 33 shows that the compacton as an internal wave dissipates energy and ultimately takes the form of a decaying oscillatory wave. Plots similar to those in FIGS. 32 and 33 for $\eta = 0.5$ are presented in FIGS. 34 and 35. The phase trajectory in FIG. 34 is almost identical to that in FIG.32. The internal oscillatory wave in FIG. 35 is also similar to the wave in FIG.33 with the only difference that the latter decays very fast.

(c) m=4 : It is evidenced by the curves in FIGS. 11 and 13 that in the nondissipative case the anticompacton solutions for equations with m=4 and m=2 are almost identical . In view of this we solved the coupled dissipative equations for m=4 with the same initial conditions as used for the case study in (a). The appropriate phase diagrams and plots of u as a function of $\xi$ for $\eta = 0.1$ and $\eta = 0.5$ are presented in FIGS.36 -39.

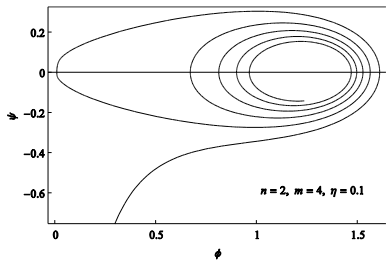

FIG.36. Phase diagram for the dissipative equation with n=2 and m=4 for $\eta = 0.1$

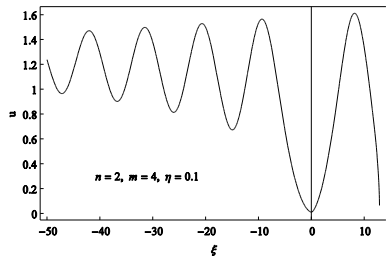

FIG.37. Solution $u$ of the dissipative equation with n=2 and m=4 for $\eta = 0.1$ as a function of $\xi$

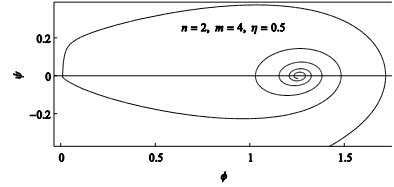

FIG.38. Same as in FIG.36 but for $\eta = 0.5$

The curves in these figures are almost identical to those presented in FIGS. 28 -31. We have verified that this is true for all equations with $m = 2k, k = 1,2,3,...$ As with the decaying solutions presented in FIGS.33 and 35 we note that the results desplayed in FIGS. 37 and 39 are also insesitive to the incressing effect of nonliearity.

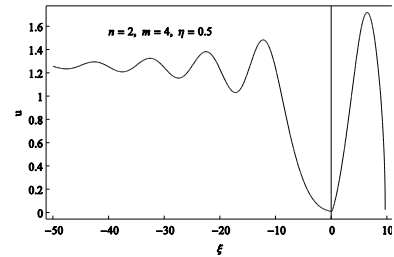

FIG.39. Same as in FIG.37 for $\eta = 0.5$

## 4. Conclusion

In addition to simulation of unidirectional propagation of small-amplitude long waves on the surface of water, the BBM equations in (3) and (5) can be used to model a wide variety of physical phenomena that arise in plasma physics, stratified fluid flows, and quantum fluid dynamics. Consequently, there exists a vast amount of literature to study the properties of these RLW equations including their group classification and connection with integrable Riccati and Abel equations [6, 13].

In this paper we made use of certain elementary concepts from the dynamical systems theory to examine how the traveling wave solutions of (3) and (5) behave as the nonlinearity and nature of dispersion in the medium modeled by them change. Equation (3) is used to study the propagation of waves in inviscid fluid while (5) refers to a dissipative BBM like system in a viscous medium.

We constructed the solutions of both non-dissipative and dissipative BBM-like equations in the traveling coordinate $\xi$. In this coordinate the partial differential equations in (3) and (5) reduce to ordinary differential equations with $\xi$ as the independent variable. Thus replacing the pair $(x,t)$ by a single variable $\xi$ we effectively make a transition from field theory to point or classical mechanics. Consequently, the variable $\xi$ may be regarded to play the same role as that of time $t$ in Newtonian mechanics.

On a very general ground one knows that in the absence of dissipation Newtonian systems are invariant under time reversal. This means that if $z(t)$ is a solution of the equation of motion, then $z(-t)$ is also a possible solution. The presence of dissipation, however, leads to violation of this discrete symmetry. It is easy to verify that the ordinary differential equations following from (3) for both n=1 and n=2 are invariant under reversal of $\xi$. As a result for every solution presented in sec. 2, we find $\phi(\xi) = \phi(-\xi)$. With regard to parity operation, $\phi \to -\phi$, we observe certain differences between linearly dispersive (n=1) and nonlinearly dispersive (n=2) equations. For instance, equations of even m for n=1 are not invariant under parity operation while the corresponding equations of odd m are found to conserve parity. On the other hand, irrespective of whether m is odd or even, all equations for n=2 are not invariant under the operation. $\phi \to -\phi$. These facts appear to have some radical effects on the phase portraits and phase trajectories of the equations.

For n=1 equations of even m have two stable points and phase trajectories are not symmetrical about the $\psi$ axis. In contrast to this, equations of odd m possess three equilibrium points and phase trajectories exhibit invariance under parity operation. For n=2, each equation of even m is characterized by one center type equilibrium point and the phase trajectory lies on the right of the $\psi$ axis resulting in the violation of reflection symmetry about the $\psi$ axis. Every odd m equation has two equilibrium points – one center lying on the right of $\psi$ axis and the other saddle situated on the left of the line $\psi = 0$. Consequently, as in the case of even m, phase trajectories of odd m equations also violate the reflection symmetry.

Although the equations of even m and of odd m for n=1 exhibit anomalous behavior with respect to their phase-space structure, all of them support soliton solutions. It may be of some interest to see how do the conserved quantities of (3) change as m increases. The conservation law for any nonlinear evolution equation in the (1+1) dimensions can be written as $T_t + X_x = 0$ in which $T$ is called the conserved density and $X$ is called the conserved flux. Admittedly, the quantity $P = \int_{-\infty}^{\infty} T dx$ is a constant of the motion.

Olver [14] in 1979 showed that the nonintegrable BBM equation (Equation (3) for m=2) has only three nontrivial conservation laws with the first, second and third conserved densities given by $T_1 = u$, $T_2 = \frac{1}{2}(u^2 + u_x^2)$ and $T_3 = \frac{1}{3}u^3$ respectively. We have verified that these are also the conserved densities for the general equation in (3) for n=1. In view of this we have calculated constants of the motion $P_1, P_2$ and $P_3$ corresponding to the conserved densities $T_1, T_2$ and $T_3$ for equations having different m values. We found that $P_1$ decreases continuously as m increases. Contrarily, $P_2$ and $P_3$ first increase as we go from m=2 to m=3 and then decrease continuously with increasing values of m. In the context of water waves, the conservation laws found by Olver are the equivalents of mass, momentum and energy conservation [15]. Thus our observed behavior of $P_1, P_2$ and $P_3$ provides us with a demonstration for the effect of nonlinearity on the conservation laws of physical systems modeled by linearly dispersive BBM like equations.

The solutions of dissipative BBM-like equations appear to exhibit some physically interesting features. For example, due to dissipative effects the soliton solutions of (3) for even values of m are transformed into undular bores while those for odd m values resemble the shock waves. With regard to the solutions of nonlinear dispersive equations we note that, as in the case of solitons of even m equations, the anticompactons are transformed to undular bores by the effects of dissipation. On the other hand, due to dissipative effects the compacton tends to vanish like the solution of an over-damped harmonic oscillator.

**Acknowledgement**

The authors would like to thank Drs. Sk. Golam Ali and Amitava Choudhuri for their kind interest in this work.